\documentclass[aps,prb,twocolumn,showpacs,amsmath]{revtex4}
\usepackage{latexsym}
\usepackage{amssymb}
\usepackage{bm}
\usepackage{graphicx}

\begin{document}

\title{Thermal rectification effect of an interacting quantum dot}

\author{CHEN Xue-Ou}
\affiliation{Department of Physics, Shanghai Jiaotong University, 1954 Huashan Road, Shanghai 200030, China \\
    Tel:021-34200557(home), 021-54743591(office), 13916453220, Email: hicxo@sjtu.edu.cn}

\author{DONG Bing}
\affiliation{Department of Physics, Shanghai Jiaotong University,
1954 Huashan Road, Shanghai 200030, China}

\author{LEI Xiao-Lin}
\affiliation{Department of Physics, Shanghai Jiaotong University,
1954 Huashan Road, Shanghai 200030, China}

\begin{abstract}

    We investigate nonlinear thermal transport properties of a single interacting quantum dot
    with two energy levels tunnel-coupled to two electrodes
    using nonequilibrium Green function method and Hartree-Fock decoupling
    approximation. In the case of asymmetric tunnel-couplings to two
    electrodes, for example, when the upper level of the quantum dot is open for transport,
    whereas the lower level is blocked, our calculations predict a
    strong asymmetry for the heat (energy) current, which shows that, the
    quantum dot system may act as a thermal rectifier in this specific
    situation.

\end{abstract}


\maketitle


Thermal transport properties of various nanoscale devices, e.g., a semiconductor quantum dot (QD) 
and a molecular junction, have recently gained considerable attention.\cite{Giazotto} Most 
theoretical work on thermoelectric effect of a QD attached to external leads have so far focused 
on the linear thermopower in sequential\cite{Beenakker} and cotunneling regimes,\cite{Kubala} and 
even in the strongly correlated Kondo-type transport regime.\cite{Boese,Kim} In addition, quantum 
interference effect in the thermopower has also been studies in a QD-ring 
geometry.\cite{Kim,Blanter}
A great amount of investigations of heat conduction through molecular chains connecting two 
electrodes at different temperature have been devoted to building a thermal rectifier, an analog 
of charge current rectifier first proposed by Aviram and Ratner in molecular 
transistor,\cite{aviram1974mr} which transport heat current (phononic current there) efficiently 
in one direction of the applied temperature difference between two electrodes, but block it in 
the reverse direction.\cite{Terraneo}
    The main ideal to raise a charge or thermal rectifier there is to make asymmetrical couplings
    between the device and electrodes, which is viable due to advances
    in the fabrication of nanostructures.

A very recent experiment by R. Scheibner {\it et.al.} has reported that a semiconductor QD with 
two energy levels can also behave as a thermal rectifier under some 
conditions.\cite{scheibner:qdt} To achieve asymmetric tunnel-couplings to two electrodes, they 
tuned high orbital momentum states of the QD as the active transport channel and applied high 
in-plane magnetic fields to quench the channel. This experimental results are quite interacting 
and very useful in thermoelectric application since the predicted rectifier is performed only in 
the electronic system and without the need for coupling to the phonon system. Stimulated by this 
experiment, we theoretically study, in this paper, the nonlinear heat transport through an 
interacting QD with two energy level asymmetrically coupling to two electrodes when a temperature 
difference is applied between the two electrodes, and examine in detail the reason and condition 
of appearance of thermal rectification effect in this system.

    The rest part of this paper is as follows. In
    Sec.~\ref{Sec_Model}, we introduce the model of a single QD with two energy levels and give 
the main theoretical formulation employed in this paper. In Sec.~\ref{Sec_Result}, we carry out 
numerical calculations and some detailed analyses with
    different coupling cases. Finally, a brief summary is given in Sec.~\ref{Sec_Conclusions}.

\begin{figure}[htb]
        \includegraphics[height=3.5cm,width=8cm]{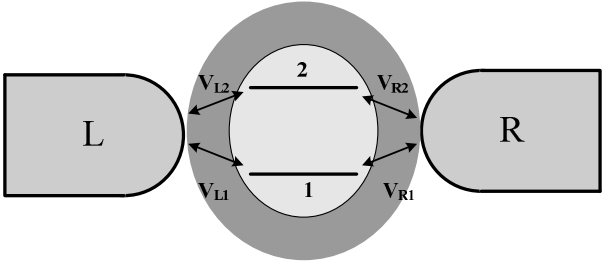}
        \caption{Schematic diagram for a single QD
            with two energy levels coupling to two electrodes.}
        \label{fig_model}
    \end{figure}

    In this paper, we consider a single interacting QD with two energy levels
    tunnel-coupling to two normal leads, as
    illustrated in Fig.~\ref{fig_model}.
    The model Hamiltonian of this system can be written as:
    \begin{equation}
        H=H_{L}+H_{R}+H_{D}+H_{I}, \label{Eq_H}
    \end{equation}
    where $H_{\eta}$ ($\eta=L,R$) describes noninteracting electron baths in the left and right     
leads, respectively:
    \begin{equation}
        H_{\eta} = \sum_{{\bf k}} \varepsilon_{\eta{\bf k}} c_{\eta {\bf k}}^\dagger                    
c_{\eta{\bf k}}^{\phantom{\dagger}},
    \end{equation}
    with $c_{\eta{\bf k}}^\dagger$ ($c_{\eta{\bf k}}$) being the creation (annihilation) operator 
of an electron
    in lead $\eta$ with momentum ${\bf k}$, and energy $\varepsilon_{\eta {\bf k}}$
    measured from the Fermi energy of the electrode.
    The chemical potentials of the left and the right leads $\mu_{L/R}$
    are associated with the bias voltage $e\Delta V = \mu_{L} - \mu_{R}$,
    which is applied symmetrically between
    the two electrodes, i.e., $\mu_{L}=-\mu_{R}=e\Delta V/2$.
    Similarly, we assume that a temperature difference
    $\Delta T$ is also applied symmetrically between two leads, $T_{L}=T+\Delta T/2$ and
    $T_{R}=T-\Delta T/2$ ($T$ is the initial temperature of both
    electrodes). In equilibrium, we have $\Delta T=\Delta V=0$.
    $H_{D}$ describes the QD with two energy levels:
    \begin{equation}
        H_{D}=\sum_{j=1,2} \varepsilon_{j}c_{j}^\dagger c_{j}^{\phantom{\dagger}} + U c_{1}^\dagger                     
c_{1}^{\phantom{\dagger}} c_{2}^\dagger c_{2}^{\phantom{\dagger}} , \label{Eq_HD}
    \end{equation}
    where $c_{j}^\dagger$ ($c_{j}$) is the annihilation operator for a spinless electron in the     
$j$th level ($j=1,2$), with $\varepsilon_j$ being the corresponding level energy. The second term 
represents the Coulomb interaction $U$ between two energy levels.
    The final term in Eq. (\ref{Eq_H}), $H_I$, describes the tunnel-coupling between the levels 
and the electrodes:
    \begin{eqnarray}
        H_I &=& \sum_{\eta {\bf k} j} V_{\eta j} c_{\eta{\bf k}}^{\dagger }c_{j} + {\rm                     
{H.c.}}. \label{Eq_HI}
    \end{eqnarray}
    Here, we ignore spin degree of freedom for simplicity.
    Throughout, we will use units with $\hbar=k_B=e=1$.

    $\\$

    We consider a nonlinear system with a finite temperature difference $\Delta
    T$ and a bias voltage $\Delta V$ applied between the two leads, which will arouse electronic
    current $J^c$ and heat current $J^h$ flowing through the QD.
    Applying the Keldysh's nonequilibrium Green function formalism,
    one can obtain: \cite{haug2008qkt}

    \begin{eqnarray}
        J^c &=& \frac {1}{2 \pi} \int d \omega ~T(\omega)~[f_L(\omega) -             
f_R(\omega)],\label{Eq_Jc}\\
        J^h &=& \frac {1}{2 \pi} \int d \omega (\omega - \Delta V)~T(\omega)~[f_L(\omega) -
            f_R(\omega)],\label{Eq_Jh}
    \end{eqnarray}
    where $f_{\eta}(\omega)=(e^{(\omega-\mu_{\eta})/T_{\eta}}+1)^{-1}$ is the Fermi
    distribution of lead $\eta$.
    The transmission coefficient, $T(\omega)$, can be expressed using the Fourier transformation
    of the retarded and advanced Green functions of the QD, $[\bm G^{r(a)}(t,t^{\prime})]_{jj'} =
    \mp i\theta(\pm t \mp t^{\prime})\langle\{c_j^{{\phantom{\dagger}}}(t),c_{j'}^{\dag}(t^{\prime})\}\rangle$, 
as:
    \begin{eqnarray}
T(\omega)=\text {Tr} \left \{ \frac{\bm \Gamma_L \bm \Gamma_R }{\bm \Gamma_L + \bm \Gamma_R} 
\left [ \bm G^r(\omega) - \bm G^a(\omega) \right ] \right \}, \label{Eq_T}
    \end{eqnarray}
    with the constant level-width function $[\bm \Gamma]_{\eta jj'} = 2 \pi \varrho_{\eta} 
V_{\eta j} V_{\eta j'}$ ($\varrho_{\eta}$ is the density of states of lead $\eta$). So 
$\Gamma_{\eta jj}$ is the bandwidth
    of the $j$th level of the QD due to the coupling with
    lead $\eta$.
    The retarded Green function can be obtained using the standard equation of motion method with 
help of a decoupling approximation. In this paper, a Hartree-Fock decoupling scheme is employed 
and is sufficient for the present purpose since we take no account of the Kondo effect and assume 
a higher temperature in our calculations, $T\gg T_{\text K}$ ($T_{\text K}$ is the Kondo 
temperature). For example, we derive the diagonal retarded Green function as:
    \begin{eqnarray}
        G^{r}_{jj}{(\omega)} &=& \frac {\omega - \varepsilon_j - 
U(1-n_{\overline{j}})}{\Pi(\omega)},\label{Eq_Gr11}\\
        \Pi(\omega) &=& (\omega - \varepsilon_j)(\omega - \varepsilon_j - U)\cr
            &&- \Sigma^r_{jj}[\omega - \varepsilon_j - U(1-n_{\overline{j}})], \label{selfen}
    \end{eqnarray}
    with $\overline{j}=1(2)$ if $j=2(1)$.
    $\Sigma^r_{jj} = -(i/2)(\Gamma_{Lj} + \Gamma_{Rj})$ is
    the retarded self energy of the QD due to coupling to the electrodes, and $n_j = \langle 
c_j^{\dagger} c_j^{{\phantom{\dagger}}}\rangle$ is the occupation number of
    electrons in the $j$th level, which can be calculated self-consistently through the lesser 
Green function, $[\bm G^{<}(t,t^{\prime})]_{jj'} =
    i\langle c_{j'}^{\dag}(t^{\prime}) c_j^{{\phantom{\dagger}}}(t) \rangle$, as
    \begin{equation}
    n_j=-\frac{i}{2\pi} \int d\omega~ G_{jj}^<(\omega). \label{occupy}
    \end{equation}
The lesser Green function is related to the retarded Green function by the Keldysh function
    $\bm G^<=\bm G^r \bm \Sigma^< \bm G^a$ and $\Sigma_{jj'}^<=i [\Gamma_{L j} f_{L}(\omega)+ 
\Gamma_{R j} f_R(\omega) ]\delta_{jj'}$.

    Based on these equations (\ref{Eq_Jc})-(\ref{occupy}), we can study the thermoelectric
    properties of the QD in the two-terminal configuration, i.e., the heat transport through the 
QD when a temperature difference $\Delta T$ is applied between the two electrodes but the charge 
transport is vanishing due to an open circuit condition.
This external applied finite temperature difference will cause diffusion of electrons cross the 
device from one terminal to the other and generate accumulation of electrons at one of the 
terminals under the condition of open circuit. Such accumulation gradually builds a nonzero 
bias-voltage across the QD to counteract the diffusion of electrons until a steady state is 
finally reached. Numerically, this temperature-difference-induced bias voltage must be obtained 
by solving the equation
\begin{equation}
J^c=0. \label{j=0}
\end{equation}
The current formula, Eq.~(\ref{Eq_Jc}), is relevant to the electron occupation numbers, $n_1$ and 
$n_2$, through the Green functions, Eqs.~(\ref{Eq_Gr11}) and (\ref{selfen}). Meanwhile, the 
electron occupation numbers are reversely dependent on the temperature difference and induced 
bias voltage through Eq.~(\ref{occupy}). Therefore, for a given temperature difference, 
$\bigtriangleup T$, we have to solve both Eqs.~(\ref{occupy}) and (\ref{j=0}) simultaneously to 
obtain a self-consistent result for the induced bias voltage, $\bigtriangleup V$, and then use it 
in Eq.~(\ref{Eq_Jh}) to calculate the heat current through the QD.


Based on the scheme indicated above, we investigate the nonlinear thermoelectric transport 
through an interacting QD with two energy levels to search for the possible thermal rectification 
mechanism. As motivated by the recent experimental results,\cite{scheibner:qdt} we focus our 
studies on the two cases of system: the symmetric tunnel-coupling case, 
$\Gamma_{Rj}=\Gamma_{Lj}$, and the asymmetric case with a blocked lower level, $\Gamma_{R1}=0$. 
In the following numerical calculations, we choose $\Gamma_{L1}=\Gamma_{L2}=\Gamma$ as the energy 
unit. We also set the temperature of equilibrium situation as $T=4.0\, \Gamma$, the inter-level 
Coulomb interaction as $U=100\, \Gamma$, and fix the energy gap between the two levels as 
$\varepsilon_2-\varepsilon_1=10\, \Gamma$.

\begin{figure}[htb]
        \includegraphics[height=6cm,width=6cm]{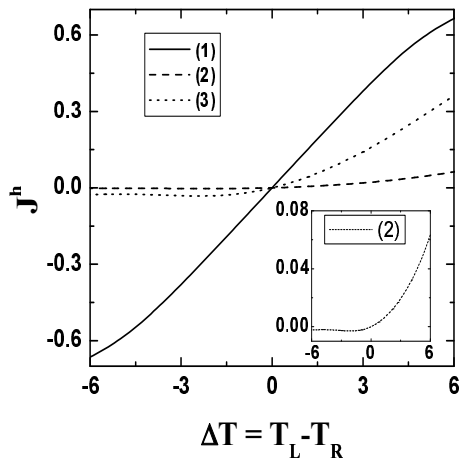}
        \caption{Heat Current vs. temperature difference with the system parameters 
$\varepsilon_1 = -6$, $\varepsilon_2 = 4$,
        for symmetric coupling case (1) $\Gamma_{R1} = 1$, $\Gamma_{R2} =1$,
        asymmetric cases (2) $\Gamma_{R1} = 0$, $\Gamma_{R2}= 1$, and
        (3) $\Gamma_{R1} = 0$, $\Gamma_{R2} = 10$. Inset: rescaling plot for the asymmetric case 
(2). The other parameters are $T=4$ and $U = 100$. All are in unit of $\Gamma$.}
        \label{fig_Jh}
    \end{figure}

\begin{figure}[htb]
        \includegraphics[height=3.8cm,width=8.5cm]{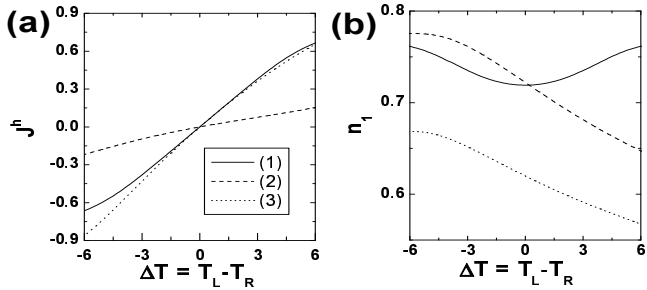}
        \caption{Heat Current $J^h$, electron occupation number of the lower level $n_1$ vs. 
temperature difference
        for symmetric coupling case (1) $\Gamma_{R1} = 1$, $\Gamma_{R2} =1$,
        upper state blocked case (2) $\Gamma_{R1} = 1$, $\Gamma_{R2}= 0$, and
        (3) $\Gamma_{R1} = 10$, $\Gamma_{R2} = 0$. Other parameters are the same as in Fig.~2.}
        \label{fig_UpperBlocking}
    \end{figure}

We show our main results in Fig.~\ref{fig_Jh}, the calculated nonlinear heat current $J^h$ as a 
function of temperature difference $\Delta T$ with different couplings
    between the two levels and electrodes.
    It is natural that for the symmetric coupling system (1), the heat current exhibits symmetric 
behavior, i.e., monotonously increases with increasing the temperature difference in both 
directions, $\Delta T>0$ and $\Delta T <0$. In contrast, a strong asymmetric behavior is found 
for the asymmetric systems (2) and (3), whose heat current increase with the temperature 
difference only in the positive direction ($\Delta T>0$ direction), but is blocked in the 
opposite direction ($\Delta T<0$), indicating the occurrence of a thermal rectification effect. 
It seems that this thermal rectification effect is a direct result of the asymmetric 
tunnel-coupling arrangement of the QD. However, the situation is not so simple because our 
calculations exhibit no thermal rectification effect for the system with blocking the higher 
state instead of the lower state [see Fig.~\ref{fig_UpperBlocking}(a) below]. Recently, an 
experimental observation of thermal rectification effect of a semiconductor QD has been reported 
by R. Scheibner, {\it et.al.}.\cite{scheibner:qdt} In their measurement, they set up some high 
orbital momentum states of the QD to play an important role in heat transport and applied high 
in-plane magnetic field to control the coupling strength between these states and the electrodes. 
They measured the gate-voltage-dependent thermopower of the QD and found that it changes to an 
asymmetric line shape when the lower state is blocked by applying a high magnetic field, which 
reveals the thermal rectification effect.
Our calculation results are in agreement with this experimental inferences.

\begin{figure}[htb]
        \includegraphics[height=4cm,width=8.5cm]{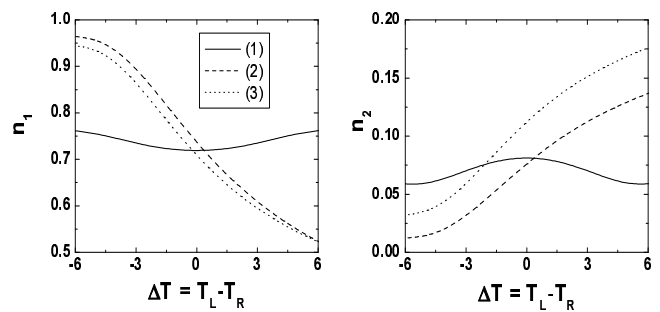}
        \caption{Electron occupation numbers vs. temperature difference
            with the same system parameters as in Fig.~\ref{fig_Jh}.}
        \label{fig_N1N2}
    \end{figure}

In order to interpret this effect, we plot the self-consistently calculated electron occupation 
numbers, $n_1$ and $n_2$, in the two levels as functions of the applied temperature difference in 
Fig.~\ref{fig_N1N2}.
It is visible that $n_1$($n_2$) of the symmetric system always increases (decreases) with 
increasing the temperature-difference in both directions, and the total electron number $n_1+n_2$ 
also increases. It means that with increasing temperature difference (in both directions), more 
and more electrons participate in the energy transferring processes, thus resulting in a 
monotonous rise of the heat current. If the lower state is blocked, we find that to fulfill the 
open circuit condition, the induced bias-voltage $\Delta V$ has a negative slope with respect to 
$\Delta T$ for the specific system parameters (not shown here).
Therefore, in the region of $\Delta T > 0$, the negative $\Delta V$ indicates a lowering chemical 
potential of the left lead $\mu_L$ with a heightening temperature $T_L$, implying a lesser 
occupation possibility $f_L(\varepsilon_1)$ for electrons in the left lead to enter the QD,
finally resulting in a lesser $n_1$ and corresponding a larger $n_2$ due to the strong Coulomb 
correlation.
In contrast, in the region of $\Delta T < 0$, a larger Fermi distribution $f_L(\varepsilon_1)$ 
makes $n_1$ increase more quickly than that of the symmetric system with $\Delta T$ arising, and 
$n_2$ decrease nearly towards $0$.
Since the upper level (level-2) is the only open transmitting level in the asymmetric system, it 
is natural that the large $n_2$ in the positive direction of $\Delta T$ leads to a large heat 
current, while the vanishing $n_2$ in the negative direction is responsible for a suppressed 
electron transfer, i.e., indicating the occurrence of a thermal rectification.

    However, the situation is different for the upper state blocked case ($\Gamma_{R2}= 0$) as 
shown in Fig.~\ref{fig_UpperBlocking}, in which the calculated heat current $J^h$, and electron 
occupation number of the lower level, $n_1$, are plotted as functions of temperature difference.
It is found that the heat current increases in both directions,
    and there occur no thermal rectification effect. This is physically because the lower 
electronic state is now the open transmitting level, thus the occupation number of this level, 
$n_1$, determines the heat current.
    From Fig.~\ref{fig_UpperBlocking}(b), it is clearly to see that
    $n_1$ keeps a relatively large value at the whole region of $\Delta T$. That is to say that 
there are sufficient numbers of electrons to carry energy transferring in both directions.

\begin{figure}[htb]
        \includegraphics[height=7cm,width=8.5cm]{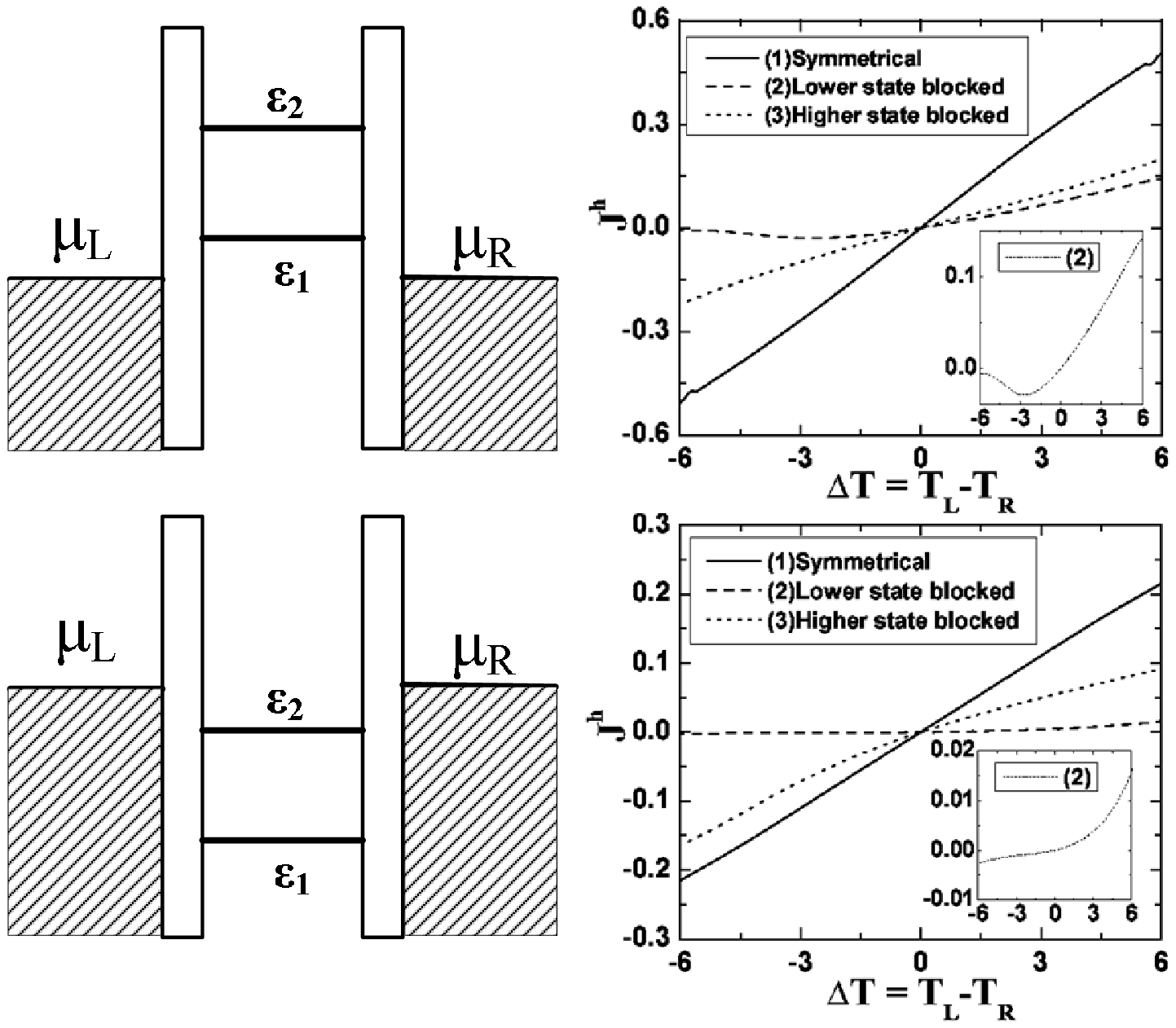}
        \caption{Heat Current $J^h$ vs. temperature difference
        for cases with two energy levels both
        above (a) or below (b) the chemical potentials of two electrodes, with parameters
        (a) $\varepsilon_1 = 4$, $\varepsilon_2 = 14$, and
        (b) $\varepsilon_1 = -16$, $\varepsilon_2 = -6$. Lines
        (1),(2),(3) are respectively corresponding to
        the symmetric coupling case (1) $\Gamma_{R1} = 1$, $\Gamma_{R2} =1$,
        lower state blocked case (2) $\Gamma_{R1} = 0$, $\Gamma_{R2}= 1$, and
        higher state blocked case (3) $\Gamma_{R1} = 1$, $\Gamma_{R2}= 0$.}
        \label{fig_UpDown}
    \end{figure}

    In above discussions, the chemical potentials of both electrodes
    are between two energy levels at equilibrium.
    One may think that it is the special arrangement of the level energies
    which decides the thermal rectification effect.
    In fact, to raise a thermal rectification effect,
    one needs to suppress the occupation number of transmitting level in one direction of $\Delta 
T$.
    For a two-level system, the lower level always holds a large occupation number.
    Thus if and only if the lower level is blocked, one can observe the thermal rectification.
    Besides, a large Coulomb interaction $U$ is also necessary for the rectification,
    because it leads to large repulsion between $n_1$ and $n_2$,
    and thus can greatly suppress the occupation number $n_2$ of the upper level.
    To verify the these conclusions, we make calculations for two cases where two energy levels
    of the QD are both above or below the chemical potentials of the two electrodes at 
equilibrium, as shown in Fig.~\ref{fig_UpDown}.
    It gives the expected result that a thermal rectification effect occurs
    for the blocking lower level cases.
    The rectification effect in Ref.~\onlinecite{scheibner:qdt} indeed corresponds to the case
    where two levels are both above the
    chemical potentials with blocking the lower state. We also calculate the nonlinear heat 
transport without the inter-level Coulomb interaction and find that there occurs no rectification 
for all symmetric and asymmetric systems (not shown here).

    In summary, we investigated the  nonlinear thermoelectric effect of a two-level quantum dot 
system
    with inter-level Coulomb interaction. We calculated the nonlinear heat current through the 
quantum dot
    when a temperature difference is applied between the two electrodes,
    and found that in some cases the heat current displays a strong asymmetrical behavior, which 
indicates a thermal rectification effect.
    We analyzed further this effect in detail,
    and ascribed it to the asymmetrical couplings between two electrodes and two levels:
    only when the upper level is open for transport and meanwhile the lower level is blocked, the 
rectification effect may occur.
    Besides, our numerical discussions shown that the strong Coulomb interaction is necessary for 
the occurrence of a rectification effect.

\begin{acknowledgments}

This work was supported by Projects of the National Science Foundation of China, the Shanghai 
Municipal Commission of Science and Technology, the Shanghai Pujiang Program, and Program for New 
Century Excellent Talents in University (NCET).

\end{acknowledgments}


\end{document}